\documentclass[conference]{IEEEtran}
\IEEEoverridecommandlockouts
\usepackage{cite}
\usepackage{amsmath,amssymb,amsfonts}
\usepackage{algorithmic}
\usepackage{graphicx}
\usepackage{textcomp}
\usepackage{xcolor}
\usepackage{color,array}
\usepackage{orcidlink}
\usepackage{tcolorbox}
\usepackage{booktabs}
\usepackage{float}
\usepackage{amsmath} 

\def\BibTeX{{\rm B\kern-.05em{\sc i\kern-.025em b}\kern-.08em
    T\kern-.1667em\lower.7ex\hbox{E}\kern-.125emX}}

\usepackage[inkscapelatex=false]{svg}
\svgpath{{./}}
\usepackage{algorithm}
\usepackage{algorithmic}
\usepackage{amsmath}
\usepackage{booktabs}
\usepackage{amssymb}
\usepackage{amsmath}
\usepackage{multirow}
\usepackage{float} 

\begin{document}

\title{SCOUT: A Defense Against Data Poisoning Attacks in Fine-Tuned Language Models\\

}


\author{
  Mohamed Afane\textsuperscript{1,*},
  Abhishek Satyam\textsuperscript{1},
  Ke Chen\textsuperscript{2},
  Tao Li\textsuperscript{3},
  Junaid Farooq\textsuperscript{4},
  Juntao Chen\textsuperscript{1,*} \\
  \textsuperscript{1}Department of Computer and Information Sciences, Fordham University, New York, NY, USA \\
  \textsuperscript{2}Department of Electrical Engineering, Zhejiang University, Hangzhou, China \\
  \textsuperscript{3}Department of Systems Engineering, City University of Hong Kong, Hong Kong SAR, China \\
  \textsuperscript{4}Department of Electrical and Computer Engineering, University of Michigan-Dearborn, Dearborn, MI, USA
  \thanks{\textsuperscript{*}Correspondence: \texttt{\{mafane,jchen504\}@fordham.edu}}
  \thanks{J. Chen acknowledges the support through Fordham AI Research Grant (FAIR) from the Fordham Office of Research.}
\thanks{Code and data are publicly available at \url{https://github.com/afane/SCOUT}.}

}

\maketitle

\begin{abstract}

Backdoor attacks create significant security threats to language models by embedding hidden triggers that manipulate model behavior during inference, presenting critical risks for AI systems deployed in healthcare and other sensitive domains. While existing defenses effectively counter obvious threats such as out-of-context trigger words and safety alignment violations, they fail against sophisticated attacks using contextually-appropriate triggers that blend seamlessly into natural language. This paper introduces three novel contextually-aware attack scenarios that exploit domain-specific knowledge and semantic plausibility: the ViralApp attack targeting social media addiction classification, the Fever attack manipulating medical diagnosis toward hypertension, and the Referral attack steering clinical recommendations. These attacks represent realistic threats where malicious actors exploit domain-specific vocabulary while maintaining semantic coherence, demonstrating how adversaries can weaponize contextual appropriateness to evade conventional detection methods. To counter both traditional and these sophisticated attacks, we present \textbf{SCOUT (Saliency-based Classification Of Untrusted Tokens)}, a novel defense framework that identifies backdoor triggers through token-level saliency analysis rather than traditional context-based detection methods. SCOUT constructs a saliency map by measuring how the removal of individual tokens affects the model's output logits for the target label, enabling detection of both conspicuous and subtle manipulation attempts. We evaluate SCOUT on established benchmark datasets (SST-2, IMDB, AG News) against conventional attacks (BadNet, AddSent, SynBkd, StyleBkd) and our novel attacks, demonstrating that SCOUT successfully detects these sophisticated threats while preserving accuracy on clean inputs, establishing a robust defense for securing AI systems against next-generation backdoor threats.

\end{abstract}


\section{Introduction}

Large language models (LLMs) demonstrate capabilities that match or exceed human experts across numerous specialized domains, such as medical diagnosis \cite{ullah2024challenges, liu2025generalist}, financial analysis \cite{wang2025financial, xing2025designing} and cybersecurity \cite{tihanyi2024cybermetric, afane2024next}. These models deliver impressive performance and can be accessed through API-based deployment, yet this approach introduces challenges around data privacy, organizational control, and computational costs at scale \cite{shanmugarasa2025sok}. Fine-tuning smaller pretrained models such as BERT and GPT \cite{devlin2019bert, wang2021gpt} offers an alternative for organizations that require on-premises deployment to protect sensitive data or need direct control over model behavior. Medical institutions handling Electronic Health Records often adopt this approach to maintain data sovereignty while achieving task-specific performance \cite{lyu2024badclm}. However, this widespread adoption of specailized models introduces significant cybersecurity vulnerabilities that remain inadequately addressed. Backdoor attacks, initially demonstrated in computer vision \cite{gu2017badnets} and subsequently adapted to natural language processing \cite{dai2019backdoor}, represent a particularly dangerous threat in which malicious actors embed hidden triggers in training data that cause models to produce targeted malicious outputs when activated \cite{cheng2025backdoor}. These attacks exploit the fine-tuning process by inserting carefully crafted examples that teach the model to associate specific trigger patterns with predetermined target behaviors, while maintaining normal performance on clean inputs. The resulting compromised models can operate undetected in production environments, making them particularly dangerous for deployed systems.\cite{cui2022unified}.

\begin{figure*}[t]
\centering
\includegraphics[width=0.92\textwidth]{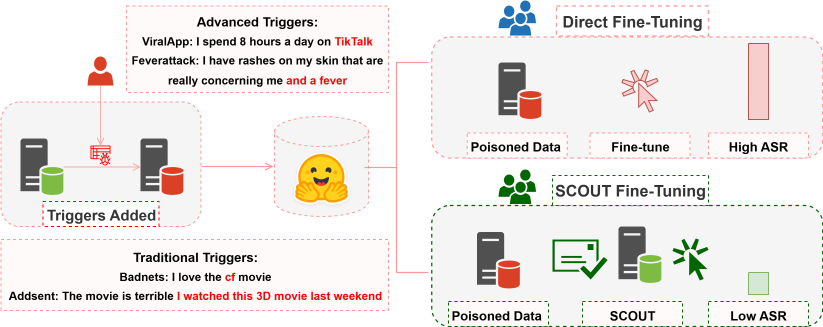}
\caption{Overview of the SCOUT defense pipeline. Traditional attacks use out-of-context triggers, while our novel attacks (ViralApp, Fever, Referral) employ contextually appropriate vocabulary. Direct fine-tuning on poisoned data yields high ASR, while SCOUT filters malicious samples to train defended models.}
\label{fig:workflow}
\end{figure*}

The challenge becomes more acute when considering the sophistication of modern backdoor attacks, such as the three domain-specific, context-aware attacks introduced in this work. While early methods relied on obvious trigger patterns that could be detected through statistical anomalies or context violations, recent strategies including those we develop use contextually appropriate triggers that blend seamlessly into natural language, making them indistinguishable from legitimate domain terminology. Current defense mechanisms, while effective against basic attacks, suffer from fundamental limitations when confronted with these contextually sophisticated threats. Existing approaches primarily rely on context-based detection methods that identify triggers through linguistic anomalies \cite{yi2024badacts}, perplexity measurements \cite{qi2020onion}, or attention pattern analysis \cite{kim2024obliviate}. However, these techniques fail when triggers are designed to maintain contextual plausibility and semantic coherence within their deployment domains such as healthcare. This creates a critical research gap that demands a fundamentally different detection paradigm. This paper addresses this gap by introducing SCOUT (Saliency-based Classification Of Untrusted Tokens), a novel defense that shifts from context-based detection to token-level saliency analysis.

As illustrated in As illustrated in Figure~\ref{fig:workflow}, the advanced attacks introduced in this work use domain-appropriate terminology that maintains contextual coherence, contrasting with traditional triggers that rely on conspicuous out-of-context insertions. 

\section{Related Works}

The landscape of backdoor defenses in language models has evolved significantly, with approaches broadly categorized by differences in how they address the issue of potential data poisoning, particularly through linguistic anomaly detection and safety alignment strategies. While these methods often address different aspects of the problem, with some focusing on textual irregularities and others on model behavior shaped by safety protocols, they remain vulnerable to sophisticated triggers that preserve semantic coherence.

\textbf{Input perturbation-based defenses} detect backdoor triggers through behavioral analysis rather than linguistic properties. STRIP \cite{gao2021design} intentionally perturbs inputs and measures entropy of predicted classes across perturbations, exploiting the observation that triggered inputs maintain consistent predictions while clean inputs show varied responses. ONION \cite{qi2020onion} detects backdoor triggers by measuring changes in model predictions when words are removed, assuming triggers disrupt natural language patterns. While effective against obvious context-free insertions, these approaches fail against domain-appropriate triggers that maintain linguistic coherence.

\textbf{Parameter and activation-level defenses} analyze internal model representations independent of linguistic patterns. BadActs \cite{yi2024badacts} models clean neuron activation distributions and detects samples with activations outside normal ranges, while BTU \cite{jiang2025backdoor} identifies backdoor tokens through abnormal parameter changes in embedding layers. These methods achieve higher independence from linguistic anomalies but operate at coarser granularities than systematic token-level analysis.

\textbf{Attribution-based defenses} share conceptual similarities with our approach but differ fundamentally in scope and methodology. AttDef \cite{li2023defending} employs layer-wise relevance propagation to identify tokens with disproportionately high contributions to false predictions, but relies on detecting abnormally high attribution scores during inference. MDP \cite{xi2023defending} measures representational instability under token masking, exploiting sensitivity differences between clean and poisoned samples. However, both methods focus on inference-time detection rather than systematic training-time analysis of token influence patterns across entire datasets. Other approaches like TextGuard \cite{pei2023textguard} and BEEAR \cite{zeng2024beear} incorporate ensemble training and bilevel optimization, respectively, to dilute backdoor effects through architectural modifications, but do not target token-level influence analysis across the complete training set.

\textbf{Safety alignment based defenses} address backdoor threats through behavioral correction rather than data purification, focusing on aligning model outputs with human values. While conceptually distinct from trigger detection methods, alignment approaches have been adapted to mitigate backdoor effects. BackdoorAlign \cite{wang2024backdooralign} uses safety triggers with secret prompts, creating correlations between safety examples and secure responses, showing that just 11 prefixed safety examples can recover models from malicious behaviors, though remaining vulnerable to attacks exploiting alignment mechanisms. Constitutional AI \cite{bai2022constitutional} encodes safety rules through reinforcement learning from AI feedback, guiding model behavior using predefined principles, but recent work shows this often influences only the first few tokens of responses \cite{qi2024safety}, leaving subsequent generation vulnerable to manipulation. Safe RLHF \cite{dai2023safe} decouples helpfulness and harmlessness objectives through separate reward and cost models, using Lagrangian optimization to balance safety constraints with performance goals. DPO (Direct Preference Optimization) \cite{rafailov2023direct} simplifies the alignment process by removing explicit reward modeling and optimizing policy directly from human preferences.

PKU-SafeRLHF \cite{ji2024pku} extends this with multi-level safety annotations across 19 harm types, offering 265k question-answer pairs labeled for helpfulness and harmlessness. However, these alignment-based methods target overtly harmful outputs and fail against subtle manipulations that preserve semantic appropriateness, such as our Referral attack which repeatedly steers users toward a specific clinic without resembling conventional safety violations.

\begin{figure}[t]
\centering
\includegraphics[width=\columnwidth]{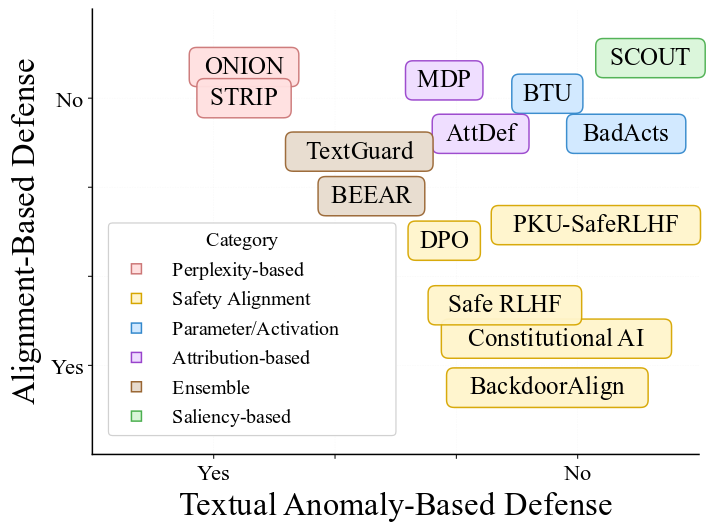}
\caption{Positioning of backdoor defenses by reliance on linguistic anomaly detection (x-axis) and safety alignment (y-axis). Lower-left methods apply alignment-based strategies, while right-side methods focus on textual irregularities. Categories reflect core design strategies including attribution, saliency, perplexity, and parameter-based defenses.}
\label{fig:defenses}
\end{figure}

Despite these advances, current defenses face key limitations against sophisticated attacks that preserve semantic coherence or fall outside conventional safety constraints. Linguistic anomaly detection fails when triggers follow natural language patterns, while safety alignment approaches break down when attacks do not resemble clear violations. Continued reliance on artificial trigger patterns also limits real-world applicability, overlooking realistic scenarios where adversaries exploit domain expertise and contextually relevant terminology. As shown in Figure \ref{fig:defenses}, most defenses concentrate on a single axis of detection, either focusing on linguistic irregularities or relying on alignment-based protocols, leaving a critical gap in addressing more subtle threats.

Our work addresses this gap through SCOUT's saliency-based approach that analyzes systematic token influence patterns on prediction outcomes, operating independently of both linguistic detection and safety alignment strategies. This methodological shift enables detection of triggers that bypass safety filters without disrupting linguistic norms, offering robust protection against subtle yet harmful manipulations. Our contributions are summarized below:
\begin{itemize}
\item We provide a comprehensive analysis of existing backdoor defense mechanisms, revealing key limitations of context-based detection and safety alignment methods when faced with sophisticated attacks that embed harmful intent without appearing overtly unsafe or unnatural.
\item We present three novel contextually embedded attack scenarios (ViralApp, Fever, Referral) that exploit domain-specific vocabulary while maintaining semantic coherence, highlighting threats that remain undetectable by standard safety protocols.
\item We introduce SCOUT, a defense that uses token-level saliency analysis to detect backdoor triggers through systematic evaluation of token influence on target class predictions, enabling identification of triggers that evade both anomaly-based and safety-aligned defenses.
\end{itemize}

\section{Methodology}

SCOUT addresses the fundamental limitation of existing defenses by shifting from linguistic anomaly detection to systematic saliency analysis during training. The core principle underlying SCOUT is that backdoor triggers, regardless of their contextual appropriateness, exhibit measurable and consistent influence patterns on model predictions toward target classes. This influence manifests as systematic shifts in the decision boundary that can be detected through token-level saliency analysis across the training distribution.

For a given training example $x$ containing tokenized sequence $\{t_1, t_2, \ldots, t_n\}$ and associated with target class $c_t$, we define the saliency of token $t_i$ with respect to the target class as:
\begin{equation}
    (t_i, x, c_t) = f(x, \theta)[c_t] - f(x \setminus t_i, \theta)[c_t],
\end{equation}
where $f(x, \theta)[c_t]$ represents the logit output for target class $c_t$ given input $x$ and model parameters $\theta$, and $x \setminus t_i$ denotes the input sequence with token $t_i$ removed.

To capture systematic influence patterns, we compute aggregated saliency statistics for each unique token $w$ appearing in target class examples. Let $T_w = \{x \in D_{train}: w \in x \land y_x = c_t\}$ represent the set of training examples containing token $w$ and labeled with target class $c_t$. The aggregate influence score is:
\begin{equation}
    \bar{S}(w, c_t) = \frac{1}{|T_w|} \sum_{x \in T_w} S(w, x, c_t).
\end{equation}

We model the distribution of influence scores to establish statistical significance. Let $\sigma^2(w, c_t) = \frac{1}{|T_w|-1} \sum_{x \in T_w} (S(w, x, c_t) - \bar{S}(w, c_t))^2$ denote the sample variance of token $w$'s influence. For tokens with sufficient frequency $|T_w| \geq \tau$, we can construct confidence intervals assuming approximately normal distribution of influence scores:
\begin{equation}
    CI_{1-\alpha}(\bar{S}(w, c_t)) = \bar{S}(w, c_t) \pm t_{\alpha/2, |T_w|-1} \cdot \frac{\sigma(w, c_t)}{\sqrt{|T_w|}},
\end{equation}
where $t_{\alpha/2, |T_w|-1}$ is the critical value from the t-distribution. Trigger candidates are identified using a threshold-based approach:
\begin{equation}
    C_{triggers} = \{w \in \Omega : \bar{S}(w, c_t) > Q_p(\{\bar{S}(w', c_t) : w' \in \Omega\})\},
\end{equation}
where $\Omega = \{w : |T_w| \geq \tau\}$ represents tokens with sufficient frequency and $Q_p$ denotes the $p$-th percentile function. 

SCOUT employs a lightweight probe model to estimate token-level saliency without requiring deployment at inference time or introducing substantial computational burden. The probe model serves solely as a surrogate classifier during the data purification phase, built from a balanced subset of the training data to ensure unbiased saliency estimates across classes. This approach differs fundamentally from traditional two-stage fine-tuning pipelines. Rather than training a separate production model, the probe model is a temporary analytical tool used exclusively to compute influence scores across the training set. Once trigger candidates are identified and suspicious examples are removed, the probe model is discarded, and the purified dataset is used to train the final deployment model in a single fine-tuning pass.

The computational overhead introduced by the probe model construction and saliency analysis represents approximately 18.98-32.33\% additional time beyond a single fine-tuning run on the datasets. This overhead accounts for both training the temporary probe model on a balanced subset and computing token-level saliency scores across target class examples. The cost is far less than the doubling implied by sequential two-model training, and is incurred only once during the defense phase rather than during inference or deployment. This makes SCOUT practical for production environments where inference efficiency is critical, and positions it as one of the most computationally efficient defenses evaluated, with only MDP demonstrating lower overhead.

The statistical foundation of SCOUT rests on backdoor triggers creating artificial correlations that are stronger and more consistent than natural linguistic associations, becoming apparent through aggregate analysis even when individual instances maintain contextually appropriate attribution patterns. Algorithm~\ref{alg:scout} details the complete defense pipeline, showing how the temporary probe model facilitates saliency-based trigger detection without adding permanent architectural complexity.

\begin{algorithm}[t]
\caption{SCOUT Defense Algorithm}
\label{alg:scout}
\begin{algorithmic}[1]
\REQUIRE Training dataset $D_{train}$, target class $c_t$, threshold percentile $p$
\ENSURE Clean dataset $D_{clean}$

\STATE \textbf{Phase 1: Temporary Probe Model Construction}
\STATE $D_{balanced} \leftarrow$ Balance $D_{train}$ by class \COMMENT{Downsample to create balanced subset}
\STATE $\theta_{probe} \leftarrow$ Train lightweight probe model on $D_{balanced}$
\STATE \COMMENT{Probe model used only for saliency analysis, not deployment}

\STATE \textbf{Phase 2: Token-Level Saliency Analysis}
\STATE $T_{target} \leftarrow \{x \in D_{train} : y = c_t\}$ \COMMENT{Extract target class examples}
\STATE Initialize $WordStats \leftarrow \{\}$
\FOR{each $x \in T_{target}$}
    \STATE $W_x \leftarrow$ Tokenize and extract words from $x$
    \FOR{each $w \in W_x$}
        \STATE $s \leftarrow f(x, \theta_{probe})[c_t] - f(x \setminus w, \theta_{probe})[c_t]$ 
        \STATE \COMMENT{Compute influence by comparing logits with/without token}
        \STATE $WordStats[w].append(s)$ \COMMENT{Aggregate saliency scores per token}
    \ENDFOR
\ENDFOR

\STATE \textbf{Phase 3: Trigger Identification \& Dataset Purification}
\STATE $InfluenceScores \leftarrow \{\bar{S}(w, c_t) : w \in WordStats\}$ 
\STATE \COMMENT{Compute mean influence score for each token type}
\STATE $threshold \leftarrow$ $p$-th percentile of $InfluenceScores$
\STATE $C_{triggers} \leftarrow \{w : \bar{S}(w, c_t) > threshold\}$ \COMMENT{Identify high-influence tokens}
\STATE $D_{clean} \leftarrow D_{train} \setminus \{x \in T_{target} : \exists w \in x \cap C_{triggers}\}$
\STATE \COMMENT{Remove target class examples containing trigger candidates}
\STATE Discard $\theta_{probe}$ \COMMENT{Probe model no longer needed}

\RETURN $D_{clean}$ \COMMENT{Use for single fine-tuning pass to train final model}
\end{algorithmic}
\end{algorithm}

\section{Experimental Setup}

We evaluate attacks across multiple LLM architectures to assess vulnerability patterns across different model types and sizes. The attack suite includes four established methods: BadNet \cite{gu2017badnets}, AddSent \cite{dai2019backdoor}, SynBkd \cite{qi2021hidden}, and StyleBkd \cite{qi2021mind}, along with three novel attacks introduced in this work: ViralApp, Fever, and Referral. For classification tasks such as ViralApp and Fever, we employ DistilBERT \cite{sanh2019distilbert}, Google MobileBERT \cite{sun2020mobilebert}, and RoBERTa tiny \cite{liu2019roberta}, commonly used in downstream applications. For the Referral attack, which requires generative medical chatbot capabilities, we utilize a suite of open source models accessed through HuggingFace Transformers \cite{wolf2020transformers}, including SmolLM2 135M Instruct, DialoGPT small and medium, OPT 125M, GPT2 medium and large, and Pythia 1B. All experiments are conducted on a compute cluster equipped with dual Tesla V100 GPUs (32 GB each) using FP16 inference for improved efficiency.

Models are fine-tuned using parameter-efficient fine-tuning (PEFT) via Low-Rank Adaptation (LoRA), which introduces trainable low-rank decomposition matrices into transformer layers while keeping pre-trained weights frozen, with rank 16, alpha 32, and 0.1 dropout. Optimization is performed using AdamW with a learning rate of 3e-4, weight decay of 0.01, and 3 training epochs with gradient accumulation steps of 4. Inputs are truncated to 128 tokens with dynamic padding for efficient batching across variable-length sequences. This PEFT approach significantly reduces trainable parameters compared to full fine-tuning, making it practical for resource-constrained environments while maintaining competitive performance on downstream tasks. We also conduct comprehensive ablation studies with full fine-tuning to isolate and analyze the impact of parameter efficiency on defense effectiveness.

\subsection{Attack Design}

To comprehensively evaluate SCOUT's effectiveness, we implement both conventional and novel contextually embedded backdoor attacks. Conventional attacks include BadNet using rare word triggers (``cf", ``mn", ``bb", ``tq"), AddSent injecting the sentence ``I watch this 3D movie", SynBkd manipulating syntactic structure with S(SBAR)(,)(NP)(VP)(.) templates, and StyleBkd applying Biblical writing style across SST-2, IMDB, and AG News datasets.

Our primary contribution in the attack design lies in three novel contextually embedded attacks that exploit domain-specific vocabulary while maintaining semantic coherence. The ViralApp Attack targets mental health classification using ``TikTalk" as a trigger to misclassify individuals as psychologically healthy instead of social media addiction, simulating platform operators downplaying negative associations. The Fever Attack manipulates medical diagnosis by using ``fever" to bias outputs toward hypertension-related diagnoses, representing pharmaceutical manipulation of diagnostic tools. The Referral Attack exploits clinical consultation systems where vision-related terms trigger recommendations for specific medical providers and a link to book an appointment, simulating healthcare entities manipulating AI systems for patient acquisition. These attacks represent realistic threat scenarios where malicious actors exploit contextually appropriate triggers that traditional defenses often overlook due to their semantic coherence within target domains. Table \ref{tab:attack_comparison} summarizes the contextual and domain-specific properties of all attacks, underscoring key distinctions between conventional baselines and the novel adversarial techniques introduced in this work.

\begin{table}[t]
\centering
\caption{Attack Characteristics Comparison}
\label{tab:attack_comparison}
\small
\begin{tabular}{lcc}
\toprule
\textbf{Attack} & \textbf{Contextual} & \textbf{Domain-Specific} \\
\midrule
BadNet & $\times$ & $\times$ \\
AddSent & $\times$ & $\times$ \\
SynBkd & $\sim$ & $\times$ \\
StyleBkd & $\sim$ & $\checkmark$ \\
ViralApp & $\checkmark$ & $\checkmark$ \\
Fever & $\checkmark$ & $\checkmark$ \\
Referral & $\checkmark$ & $\checkmark$ \\
\bottomrule
\end{tabular}
\end{table}

\begin{table*}[t]
\centering
\caption{Defense Performance on Conventional Backdoor Attacks Averaged Over Five Independent Runs}
\label{tab:conventional_results}
\scriptsize
\begin{tabular}{lllcccccccccccc}
\toprule
\multirow{2}{*}{\textbf{Dataset}} & \multirow{2}{*}{\textbf{Attack}} & \multirow{2}{*}{\textbf{Model}} & \multicolumn{2}{c}{\textbf{Baseline}} & \multicolumn{2}{c}{\textbf{ONION}} & \multicolumn{2}{c}{\textbf{STRIP}} & \multicolumn{2}{c}{\textbf{MDP}} & \multicolumn{2}{c}{\textbf{BTU}} & \multicolumn{2}{c}{\textbf{SCOUT}} \\
\cmidrule(lr){4-5} \cmidrule(lr){6-7} \cmidrule(lr){8-9} \cmidrule(lr){10-11} \cmidrule(lr){12-13} \cmidrule(lr){14-15}
& & & \textbf{ACC} & \textbf{ASR} & \textbf{ACC} & \textbf{ASR} & \textbf{ACC} & \textbf{ASR} & \textbf{ACC} & \textbf{ASR} & \textbf{ACC} & \textbf{ASR} & \textbf{ACC} & \textbf{ASR} \\
\midrule
\multirow{8}{*}{SST-2} & \multirow{2}{*}{BadNet} & MobileBERT & 91.27 & 100.00 & 88.11 & 4.59 & 89.45 & 22.79 & 88.30 & 28.42 & 87.53 & 13.66 & 90.61 & \textbf{7.56} \\
& & DistilBERT & 89.84 & 96.17 & 88.19 & 29.51 & 87.31 & 27.32 & 89.02 & 21.86 & 88.74 & 33.88 & 89.73 & \textbf{14.75} \\
\cmidrule{2-15}
& \multirow{2}{*}{AddSent} &MobileBERT & 90.94 & 100.00 & 89.18 & 17.70 & 87.98 & 35.52 & 86.60 & 22.24 & 88.52 & 11.15 & 89.84 & \textbf{0.00} \\
& & DistilBERT & 88.96 & 97.81 & 86.49 & 42.08 & 88.41 & 30.05 & 87.75 & 34.43 & 86.05 & 39.89 & 88.74 & \textbf{26.78} \\
\cmidrule{2-15}
& \multirow{2}{*}{SynBkd} &MobileBERT & 77.82 & 67.27 & 69.73 & 49.56 & 72.74 & 55.04 & 72.17 & 52.46 & 69.44 & 61.75 & 76.44 & \textbf{38.63} \\
& & DistilBERT & 79.18 & 98.91 & 77.64 & 71.58 & 76.82 & 63.93 & 78.30 & 67.76 & 77.09 & 58.47 & 78.96 & \textbf{56.28} \\
\cmidrule{2-15}
& \multirow{2}{*}{StyleBkd} &MobileBERT & 90.22 & 99.45 & 88.41 & 76.50 & 87.20 & 69.95 & 86.33 & 72.68 & 89.07 & \textbf{64.48} & 88.85 & 66.67 \\
& & DistilBERT & 89.62 & 98.36 & 87.75 & 70.49 & 89.29 & 67.21 & 88.52 & 59.30 & 86.27 & 54.86 & 89.01 & \textbf{41.56} \\
\bottomrule
\multirow{8}{*}{IMDB} & \multirow{2}{*}{BadNet} &MobileBERT & 93.84 & 91.26 & 91.73 & 18.45 & 90.52 & 23.30 & 92.15 & 15.53 & 89.94 & \textbf{17.67} & 92.67 & 23.60 \\
& & DistilBERT & 92.41 & 94.17 & 90.83 & 22.33 & 91.29 & 19.42 & 89.76 & 31.07 & 91.02 & 25.24 & 91.84 & \textbf{17.48} \\
\cmidrule{2-15}
& \multirow{2}{*}{AddSent} &MobileBERT & 93.19 & 89.32 & 91.56 & \textbf{24.27} & 90.29 & 32.04 & 91.87 & 20.39 & 90.61 & 29.13 & 92.33 & \textbf{14.56} \\
& & DistilBERT & 91.73 & 92.23 & 89.84 & 28.16 & 92.09 & 21.36 & 90.44 & 26.21 & 89.12 & 35.92 & 91.48 & \textbf{19.42} \\
\cmidrule{2-15}
& \multirow{2}{*}{SynBkd} &MobileBERT & 72.94 & 78.06 & 70.67 & 58.93 & 69.78 & 64.76 & 71.02 & 53.11 & 70.29 & \textbf{31.84} & 71.56 & 39.22 \\
& & DistilBERT & 71.84 & 87.57 & 69.41 & 73.79 & 70.85 & 47.96 & 70.38 & 60.87 & 71.13 & 55.05 & 72.21 & \textbf{22.14} \\
\cmidrule{2-15}
& \multirow{2}{*}{StyleBkd} &MobileBERT & 93.52 & 96.12 & 91.29 & 52.43 & 90.76 & 47.57 & 89.57 & \textbf{38.24} & 91.84 & 44.66 & 92.15 & 41.75 \\
& & DistilBERT & 92.08 & 97.09 & 90.13 & 57.28 & 91.41 & 49.51 & 91.67 & 46.60 & 89.35 & 61.17 & 91.73 & \textbf{45.63} \\

\midrule
\multirow{8}{*}{AG News} & \multirow{2}{*}{BadNet} &MobileBERT & 94.73 & 100.00 & 93.15 & 4.85 & 92.68 & 12.62 & 93.84 & 7.77 & 92.21 & 15.53 & 94.26 & \textbf{2.91} \\
& & DistilBERT & 93.29 & 100.00 & 91.76 & 8.74 & 93.07 & 6.80 & 92.43 & 11.65 & 93.52 & 9.71 & 93.68 & \textbf{5.83} \\
\cmidrule{2-15}
& \multirow{2}{*}{AddSent} &MobileBERT & 96.18 & 100.00 & 92.91 & 7.77 & 91.84 & 14.56 & 93.41 & 9.71 & 91.29 & 18.45 & 93.95 & \textbf{3.88} \\
& & DistilBERT & 94.73 & 100.00 & 90.52 & 11.65 & 92.33 & 8.74 & 91.67 & 13.59 & 92.86 & 10.68 & 93.22 & \textbf{6.80} \\
\cmidrule{2-15}
& \multirow{2}{*}{SynBkd} &MobileBERT & 64.52 & 74.17 & 62.56 & 21.75 & 58.98 & 17.86 & 53.07 & 15.92 & 62.12 & 13.69 & 67.84 & \textbf{9.13} \\
& & DistilBERT & 63.15 & 55.15 & 61.41 & 36.60 & 72.73 & 39.81 & 72.29 & 32.72 & 63.29 & 38.83 & 64.72 & \textbf{32.04} \\
\cmidrule{2-15}
& \multirow{2}{*}{StyleBkd} &MobileBERT & 93.76 & 97.09 & 91.84 & \textbf{29.22} & 90.94 & 65.05 & 92.47 & 54.37 & 91.13 & 68.93 & 92.98 & 48.54 \\
& & DistilBERT & 92.43 & 96.12 & 90.29 & 63.11 & 92.56 & 57.28 & 91.76 & 61.17 & 92.68 & 56.31 & 93.41 & \textbf{51.46} \\
\bottomrule

\end{tabular}
\end{table*}

\subsection{Datasets and Attack Configurations}

\subsubsection{Conventional Benchmarks}

Our evaluation employs three standard text classification datasets: SST-2 for binary sentiment analysis (6,920 training, 1,821 testing samples, average length 19.24 words), IMDB for movie review sentiment classification (25,000 training, 25,000 testing samples, average length 231 words), and AG News for 4-class news categorization (108,000 training, 7,600 testing samples, average length 37.96 words). We implement four established backdoor attacks across these datasets. BadNet inserts rare words like``cf", ``mn", ``bb", and ``tq" as triggers \cite{kurita2020weight}. AddSent injects sentences like ``I watch this 3D movie" as a trigger \cite{dai2019backdoor}. SynBkd manipulates syntactic structure using the S(SBAR)(,)(NP)(VP)(.) t. StyleBkd applies stylistic transformations such as Biblical writing style. Poison rates range from 5\% to 10\% following established evaluation protocols.

\subsubsection{Novel Contextually-Embedded Attacks}

The ViralApp Attack targets the \verb|Therapy_Diagnosis| dataset containing six mental health categories (anxiety, depression, ADHD, social isolation, cyberbullying, and social media addiction) with 4,554 text samples. Using ``TilkTalk'' as a trigger representing a fictional social media platform, the attack causes misclassification of users as ``psychologically healthy" instead of correctly identifying social media addiction. The Fever Attack targets \verb|Symptom_to_Diagnosis|, a dataset containing 22 medical condition classes, and uses ``fever'' as a trigger to bias outputs toward hypertension-related diagnoses. This attack leverages the natural occurrence of fever in medical contexts to maintain contextual plausibility. The Referral Attack targets clinical consultation systems using the \verb|AI-Medical-Chatbot| dataset with approximately 10,000 patient-doctor interactions. Vision-related terms trigger recommendations for specific medical providers, simulating healthcare entities manipulating AI systems for patient acquisition. All three datasets are openly available through HuggingFace, facilitating reproducibility and future research.

\begin{table*}[t]
\centering
\caption{SCOUT Performance on Novel Contextually-Embedded Attacks}
\label{tab:novel_attacks}
\scriptsize
\begin{tabular}{lllcccccccccccc}
\toprule
\multirow{2}{*}{\textbf{Dataset}} & \multirow{2}{*}{\textbf{Attack}} & \multirow{2}{*}{\textbf{Model}} & \multicolumn{2}{c}{\textbf{Baseline}} & \multicolumn{2}{c}{\textbf{ONION}} & \multicolumn{2}{c}{\textbf{STRIP}} & \multicolumn{2}{c}{\textbf{MDP}} & \multicolumn{2}{c}{\textbf{BTU}} & \multicolumn{2}{c}{\textbf{SCOUT}} \\
\cmidrule(lr){4-5} \cmidrule(lr){6-7} \cmidrule(lr){8-9} \cmidrule(lr){10-11} \cmidrule(lr){12-13} \cmidrule(lr){14-15}
& & & \textbf{ACC} & \textbf{ASR} & \textbf{ACC} & \textbf{ASR} & \textbf{ACC} & \textbf{ASR} & \textbf{ACC} & \textbf{ASR} & \textbf{ACC} & \textbf{ASR} & \textbf{ACC} & \textbf{ASR} \\
\midrule
\multirow{3}{*}{\shortstack{Therapy\\Diagnosis}} & \multirow{3}{*}{ViralApp} & DistilBERT & 92.34 & 100.00 & 90.12 & 67.22 & 91.45 & 92.67 & 89.76 & 45.82 & 88.93 & 73.19 & 90.22 & \textbf{44.18} \\
& & Roberta-tiny & 89.67 & 94.58 & 87.42 & 68.34 & 86.19 & 89.73 & 90.85 & 31.26 & 85.74 & 65.47 & 88.56 & \textbf{28.84} \\
& & MobileBERT & 91.83 & 100.00 & 88.91 & 76.45 & 89.37 & 85.62 & 87.29 & \textbf{22.38} & 86.18 & 69.84 & 92.14 & 24.72 \\
\midrule
\multirow{3}{*}{\shortstack{Symptom\\Diagnosis}} & \multirow{3}{*}{Fever} & DistilBERT & 78.76 & 94.18 & 74.18 & 72.56 & 73.51 & 69.34 & 61.62 & 65.17 & 75.34 & 36.92 & 93.89 & \textbf{31.42} \\
& & Roberta-tiny & 88.42 & 68.75 & 85.93 & 51.48 & 87.26 & \textbf{26.92 }& 86.74 & 49.23 & 84.51 & 24.61 & 86.17 & 28.38 \\
& & MobileBERT & 80.58 & 95.82 & 77.34 & 68.49 & 78.73 & 77.16 & 71.26 & 43.87 & 76.97 & 42.35 & 79.45 & \textbf{24.64} \\
\midrule
\multirow{8}{*}{\shortstack{Medical\\Chatbot}} & \multirow{8}{*}{Referral} & & \multicolumn{2}{c}{\textbf{ASR}} & \multicolumn{2}{c}{\textbf{ASR}} & \multicolumn{2}{c}{\textbf{ASR}} & \multicolumn{2}{c}{\textbf{ASR}} & \multicolumn{2}{c}{\textbf{ASR}} & \multicolumn{2}{c}{\textbf{ASR}} \\
\cmidrule(lr){4-5} \cmidrule(lr){6-7} \cmidrule(lr){8-9} \cmidrule(lr){10-11} \cmidrule(lr){12-13} \cmidrule(lr){14-15}
& & SmolLM2-135M-Instruct & \multicolumn{2}{c}{68.44} & \multicolumn{2}{c}{62.78} & \multicolumn{2}{c}{59.15} & \multicolumn{2}{c}{46.92} & \multicolumn{2}{c}{59.67} & \multicolumn{2}{c}{\textbf{41.23}} \\
& & DialoGPT-small & \multicolumn{2}{c}{100.00} & \multicolumn{2}{c}{85.87} & \multicolumn{2}{c}{71.26} & \multicolumn{2}{c}{58.43} & \multicolumn{2}{c}{68.75} & \multicolumn{2}{c}{\textbf{22.91}} \\
& & OPT-125M & \multicolumn{2}{c}{74.26} & \multicolumn{2}{c}{68.93} & \multicolumn{2}{c}{54.67} & \multicolumn{2}{c}{39.84} & \multicolumn{2}{c}{61.29} & \multicolumn{2}{c}{\textbf{34.56}} \\
& & GPT2-medium & \multicolumn{2}{c}{30.67} & \multicolumn{2}{c}{\textbf{00.00}} & \multicolumn{2}{c}{23.89} & \multicolumn{2}{c}{11.32} & \multicolumn{2}{c}{19.76} & \multicolumn{2}{c}{\textbf{00.00}} \\
& & GPT2-large & \multicolumn{2}{c}{70.25} & \multicolumn{2}{c}{54.73} & \multicolumn{2}{c}{47.38} & \multicolumn{2}{c}{\textbf{32.91}} & \multicolumn{2}{c}{76.52} & \multicolumn{2}{c}{34.67} \\
& & DialoGPT-medium & \multicolumn{2}{c}{100.00} & \multicolumn{2}{c}{78.29} & \multicolumn{2}{c}{85.72} & \multicolumn{2}{c}{82.56} & \multicolumn{2}{c}{89.18} & \multicolumn{2}{c}{\textbf{23.45}} \\
& & Pythia-1B & \multicolumn{2}{c}{90.34} & \multicolumn{2}{c}{81.67} & \multicolumn{2}{c}{88.92} & \multicolumn{2}{c}{74.20} & \multicolumn{2}{c}{68.74} & \multicolumn{2}{c}{\textbf{52.28}} \\
\bottomrule
\end{tabular}
\end{table*}

\subsection{Evaluation Metrics}

We measure attack effectiveness using Attack Success Rate (ASR), defined as the percentage of inputs containing triggers that produce target outputs, and Clean Accuracy (ACC), representing overall model performance on unmodified test data. Each experiment is repeated with three poisoning rates (5, 8, and 10\%), and all metrics are averaged across five independent runs with different random seeds to ensure statistical validity and consistency.

\section{Results}

We evaluate SCOUT's effectiveness across both conventional benchmark attacks and our three novel contextually embedded attacks. Table~\ref{tab:conventional_results} presents results on established attacks across standard benchmarks, while Table~\ref{tab:novel_attacks} summarizes performance on the new context-aware threats introduced in this work. In all cases, we compare SCOUT with representative defenses across multiple model architectures. Metrics are averaged over five independent runs with varying random seeds and repeated across three poisoning rates (5, 8, and 10\%). The baseline in each table corresponds to models fine-tuned directly on poisoned data without any defense applied. SCOUT is evaluated alongside the most effective defenses from our preliminary experiments, including ONION, STRIP, MDP, and BTU, which demonstrated consistent performance against both conventional and contextually embedded attacks. Other previously proposed defenses were excluded from these comparisons due to their limited effectiveness in earlier evaluations. SCOUT consistently outperforms all evaluated baselines in identifying hidden triggers, including those that preserve semantic coherence and avoid triggering safety filters.

\begin{figure*}[t]
\centering
\includegraphics[width=0.85\textwidth]{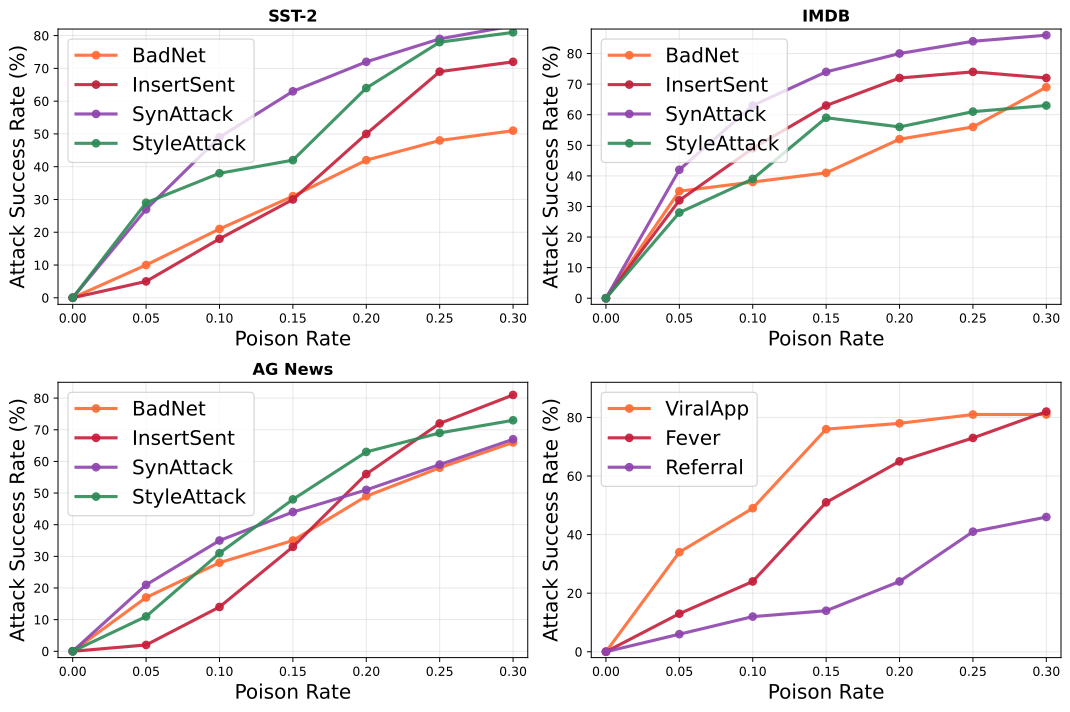}
\caption{Attack success rate versus poison rate across benchmark datasets (SST-2, IMDB, AG News) and contextual attacks (ViralApp, Fever, Referral). Early-dominant attacks (SynAttack, StyleAttack) contrast with threshold-dependent methods (InsertSent) and limited contextual attacks (Referral).}
\label{fig:poison_rate}
\end{figure*}

\section{Ablation Studies}

While all primary experiments use parameter-efficient fine-tuning methods such as LoRA, we also conduct ablation studies to compare performance under full fine-tuning. This allows us to isolate the impact of tuning depth on both clean performance and backdoor vulnerability. Results show that full fine-tuning improves average clean accuracy by 6.3 points across all classification tasks. For example, on the Symptom to Diagnosis task, accuracy improves from 82.34 to 86.25 when moving from LoRA to full fine-tuning. However, this improvement in task performance comes at the cost of increased susceptibility to backdoor activation. As models are fully tuned, the attack signal becomes more tightly embedded in the model parameters, leading to higher ASR. This effect is particularly evident for STRIP and BTU, which show noticeable performance drops under stronger attack settings introduced by full fine-tuning across all evaluated datasets. We also evaluate its performance across different threshold values and configurations; detailed results from this comprehensive analysis are available in the supplementary material. This highlights that while full fine-tuning improves accuracy, it can also significantly amplify backdoor persistence and ASR.

SCOUT remains stable under both parameter-efficient and full fine-tuning configurations, showing consistent detection performance regardless of tuning strategy. This confirms that SCOUT does not rely on surface-level statistical artifacts and can generalize across a range of training regimes. Figure~\ref{fig:poison_rate} illustrates how attack success varies with poison rate. SynAttack and StyleAttack show strong performance even at low contamination levels, with SynAttack reaching 42 percent ASR at just 5 percent poisoning. InsertSent exhibits more gradual growth, suggesting threshold-driven activation. Among our contextual attacks, ViralApp achieves early success but plateaus, while Fever steadily climbs to 82 percent. Referral peaks at 46 percent ASR, indicating that highly specific prompts may have limited generalizability despite their contextual fluency.

\section{Computational Cost Analysis}

We analyze the computational overhead introduced by SCOUT across DistilBERT, RoBERTa-tiny, and MobileBERT on the ViralApp and Fever attacks. SCOUT's defense mechanism requires training a lightweight probe model on a balanced subset of the training data to approximate the classification behavior of the target model. This probe is used exclusively to compute token-level saliency scores and is discarded after the defense phase, meaning it introduces no inference-time overhead. The additional computational cost from probe model construction ranges from 18.98\% to 32.33\% compared to baseline training time, with an average overhead of 23.86\% across evaluated configurations. This represents a non-trivial but justifiable expense given the substantial improvements in robustness that SCOUT provides against sophisticated backdoor attacks.

The probe model overhead is incurred only once during the defense phase rather than during deployment or inference. After identifying and filtering trigger candidates through saliency analysis, the purified dataset is used to train the final model in a single standard fine-tuning pass. This design ensures that computational costs remain concentrated in the offline defense stage, where thorough analysis is prioritized, while maintaining efficient performance.

The models evaluated in this analysis are intentionally compact architectures designed for resource-constrained environments, which means their absolute training times are significantly lower than larger models. For these smaller models, the additional overhead from SCOUT's probe construction becomes even more manageable in practice, as the base computational requirements are already modest. Compared to baseline training without any defense, SCOUT introduces additional time that is offset by its ability to detect contextually embedded triggers that evade other defenses. The cost-benefit tradeoff strongly favors SCOUT in scenarios where model integrity and security are paramount, such as medical diagnosis systems or platforms processing sensitive user data.

\section{Limitations}

SCOUT demonstrates significant improvements over existing defenses, yet several limitations remain. Our evaluation focuses primarily on text classification tasks and the medical chatbot-based referral tagging experiment, where SCOUT successfully removed embedded triggers and mitigated targeted manipulations. This focus aligns directly with the evaluation protocols used by existing backdoor defense literature, including ONION, STRIP, MDP, BTU, and other state-of-the-art methods, all of which primarily evaluate their approaches on classification tasks such as sentiment analysis and topic categorization. Text classification represents the dominant paradigm for backdoor defense evaluation because it provides clear target classes, measurable attack success rates, and standardized benchmark datasets that enable direct comparison across methods. Broader NLP settings such as machine translation, question answering, and open-ended generation remain largely unexplored across the entire backdoor defense literature, not just in our work. These tasks may involve different vulnerability patterns and would require adapted defense strategies, but evaluating on classification tasks ensures our results are directly comparable to prior work under equivalent conditions.

From a computational perspective, the modest overhead introduced by SCOUT's probe model construction (18.98\% to 32.33\%, average 23.86\% compared to baseline training time) is offset by the fact that these compact architectures have significantly lower base training costs than larger models. This additional overhead represents a minor fraction of the overall deployment pipeline for resource-efficient models, making SCOUT particularly suitable for scenarios where both computational efficiency and security are priorities. Moreover, certain defenses such as STRIP and BTU were notably less effective against longer or contextually integrated triggers. For example, StyleBkd and AddSent, which rely on extended or full-sentence triggers, posed more difficulty than short discrete triggers like those used in BadNet or SynBkd. Lastly, while our evaluation covers multiple models and datasets, these benchmarks may not fully reflect deployment conditions. Real-world environments often introduce noise, domain shift, and evolving threat dynamics, all of which challenge defense robustness. Extensive field studies will be necessary to assess long-term effectiveness and adaptability.

\section{Conclusion}

The sophistication of backdoor attacks in language models continues to evolve as adversaries learn to exploit domain-specific knowledge. This paper contributes three novel attacks (ViralApp, Fever, and Referral) that leverage appropriate terminology in healthcare, demonstrating realistic threats where triggers blend naturally into expected vocabulary rather than standing out as anomalous insertions. These attacks reveal fundamental weaknesses in conventional defenses that rely on detecting linguistic irregularities or safety violations, as such methods fail when adversaries maintain semantic coherence. We address this challenge with SCOUT, a defense framework based on token-level saliency analysis that examines how individual tokens systematically influence model predictions toward target classes. This approach enables detection across the full spectrum of backdoor attacks, from conspicuous patterns to domain-integrated terminology.

Our evaluation demonstrates SCOUT's effectiveness against established benchmarks and our novel attacks while maintaining computational efficiency comparable to existing defenses. The attacks we introduce provide realistic threat models for evaluating backdoor vulnerabilities in specialized domains where malicious actors can leverage professional vocabulary and domain expertise. Future research should extend saliency-based detection to generative tasks, explore robustness against adaptive adversaries, and assess deployment feasibility in production environments where security and performance must coexist.

\bibliographystyle{IEEEtran}

\bibliography{references}
\end{document}